\documentclass[aps,prc,twocolumn,showpacs,superscriptaddress,nofootinbib,longbibliography,floatfix,10pt]{revtex4-1}

\usepackage{bm}
\usepackage{dcolumn}
\usepackage{amsthm}
\usepackage{amssymb}
\usepackage{amsmath}
\usepackage{gensymb}
\usepackage{graphicx}
\usepackage{bm}
\usepackage{xcolor}
\usepackage{fix-cm}
\usepackage{mathptmx} 
\usepackage[T1]{fontenc}
\usepackage[colorlinks,allcolors=blue]{hyperref}
\makeatletter
\def\NAT@def@citea{\def\@citea{\NAT@separator}}
\makeatother

\begin{document}

\newcommand{\vcr}{\mbox{${\bf r}\,$}}

\title{Transport properties of isospin asymmetric nuclear matter using TDHF}

\author{A.S. Umar}
\affiliation{Department of Physics and Astronomy, Vanderbilt University, Nashville, Tennessee 37235, USA}
\author{C. Simenel}
\affiliation{Department of Nuclear Physics, RSPE, Australian National University, Canberra, ACT 0200, Australia}
\author{W. Ye}
\affiliation{Department of Physics, Southeast University, Nanjing 210096, Jiangsu Province, People's Republic of China}

\date{\today}

\begin{abstract}
\begin{description}
\item[Background]
The study of deep-inelastic reactions of nuclei provide a vehicle to investigate nuclear transport phenomena
for a full range of equilibration dynamics. These inquires provide us the ingredients to model such phenomena and help
answer important questions about the nuclear Equation of State (EOS) and its evolution as a function of neutron-to-proton $(N/Z)$
ratio.
\item[Purpose]
The motivation is to examine the real-time dynamics of nuclear transport phenomena and its dependence on
$(N/Z)$ asymmetry from a microscopic point of view to avoid any pre-conceived assumptions about the
involved processes. 
\item[Method]
Time-dependent Hartree-Fock (TDHF) method in full 3D is employed to calculate deep-inelastic reactions of 
$^{78}$Kr+$^{208}$Pb and $^{92}$Kr+$^{208}$Pb systems at 8.5~MeV$/A$. The impact parameter and energy-loss
dependence of relevant observables are calculated. In addition, density constrained TDHF method is used
to compute excitation energies of the primary fragments. The statistical deexcitation code GEMINI is utilized
to examine the final reaction products.
\item[Results]
The kinetic energy loss and sticking times as a function of impact parameter are calculated. 
Final properties of the fragments (charge, mass, scattering angle,
kinetic energy) are computed. Their evolution as a function of energy loss is studied and various 
intra-relations are investigated. The fragment excitation energy sharing is computed.
\item[Conclusions]
We find a smooth dependence of the energy loss, $E_\mathrm{loss}$, on the impact parameter for both
systems. On the other hand the transfer properties for low $E_\mathrm{loss}$ values are very different
for the two systems but become similar in the higher $E_\mathrm{loss}$ regime. The mean life time of the
charge equilibration process, obtained from the final $(N-Z)/A$ value of the fragments, is shown to be $\sim 0.5$~zs.
This value is slightly larger (but of the same order) than the value obtained from reactions at Fermi energies. 

\end{description}
\end{abstract}
\maketitle

\section{Introduction}
Study of strongly damped collisions of nuclei or so called deep-inelastic collisions
can play an important role in elucidating the dynamics of charge and mass exchange,
dissipation of energy and angular momentum, degree of isospin equilibration,
and the dependence of these quantities on the properties of the reactants
such as the neutron-to-proton ratio ($N/Z$)~\cite{schroder1977,moretto1981,toke1992}. 
In addition, these reactions probe an intriguing interplay between the microscopic
single-particle dynamics and collective motion at time scales too short for full
equilibration.
For these collisions Coulomb and centrifugal interactions overcome the strong
nuclear attraction and result in final fragments somewhat reminiscent  of the
initial projectile and target and thus occupy the regime between quasielastic 
and fusion/fission 
reactions.
It has also been suggested to use deep-inelastic reactions for isotope production~\cite{dasso1994}.

One of the major open questions in strongly damped reactions 
is the dependence of the final state products (and related observables) on the neutron excess,
or equivalently on the total isospin quantum number $T_z = (Z-N)/2$.
Besides being a fundamental nuclear structure and reaction question, the answer to this inquiry
is also vital to our understanding of the nuclear equation of state (EOS) and symmetry energy\,\cite{li2014}.
The EOS plays a key role in elucidating the structure of exotic nuclei\,\cite{roca2011,chen2015},
the dynamics of heavy ion collisions\,\cite{danielewicz2002,tsang2009},
the composition of neutron stars\,\cite{chamel2008},
and the mechanism 
of core-collapse supernovae\,\cite{shen2011}.

Transport properties of isospin asymmetric nuclear matter can be investigated by studying charge equilibration driven by the nuclear symmetry-energy in heavy-ion collisions.
For example, collisions at Fermi energies give access to contact times which are short enough to induce only a partial charge equilibration \cite{tsang2004} and thus can be used to determine equilibration times~\cite{jedele2017}.
Alternatively, charge equilibration has also been studied with deep inelastic collisions at lower energies, but with large isospin asymmetry in the entrance channel~\cite{planeta1988,desouza1988b,planeta1990}.
To reveal possible systematic trends requires both theoretical
and experimental studies with a wide variety of projectile and target
combinations which are expected to become available at current and future
radioactive ion-beam (RIB) facilities~\cite{balantekin2014}.
In addition, much greater degree of isospin asymmetry available with RIBs 
will allow timescale and degree of isospin equilibration to be studied in detail.

Deep-inelastic reactions at lower energies ($E/A\lesssim 20$~MeV)
 have been historically studied using statistical nucleon transport models\,\cite{ayik1976,randrup1979}. 
Many studies have concluded
that one-body dissipation~\cite{blocki1978,feldmeier1987} 
and friction models~\cite{gross1978,baldo1987}
coupled with the proper choice of collective coordinates provides
a reasonable approach to study these reactions.
However, difficulties exist in modeling these reactions due to their time-dependent non-equilibrium
nature. The portioning of excitation energy between the final
fragments~\cite{vandenbosch1984,wilczynski1989,toke1992,awes1984}, 
the amount of irreversible
energy dissipation as opposed to excitation of collective modes, conversion of angular momentum
to intrinsic spins of the final fragments~\cite{randrup1982},
influence of transfer~\cite{feldmeier1984,freiesleben1984,baldo1987} 
have all been subject to experimental~\cite{desouza1988b,planeta1989,barrett2015,rafferty2016} 
scrutiny with often less than satisfactory comparisons~\cite{schroder1984,toke1992}.
Part of the complication arises from the model dependence of the experimental analysis. For example determination of
excitation energies requires the modeling of the decay or fission of the primary fragments.
Recently, new experiments with RIBs have been proposed to elucidate some of these discrepancies~\cite{spiral-2}.
These, coupled with theoretical studies that are microscopic and dynamical in nature can further our
understanding of the dependence of these reactions on the $N/Z$ asymmetry and the shell structure of the
participating nuclei.

For these low-energy heavy-ion collisions the relative motion of the centers of the two nuclei is characterized
by a short wavelength, and thus allows for a classical treatment whereas the wavelength for the particle motion
is not small compared to nuclear sizes and should be treated quantum mechanically~\cite{umar1984a}.
The mean-field approach such as the time-dependent Hartree-Fock (TDHF) theory provides a microscopic basis for describing
heavy-ion reaction mechanism at low bombarding energies\,\cite{negele1982,simenel2012,nakatsukasa2016}.
TDHF collisions which result in well separated final fragments
provide a means to study the deep-inelastic scattering of heavy-systems,
allowing for  the calculation of
certain scattering observables, such as the final mass, charge, and
scattering angles of the fragments. These are simply calculated by
taking expectation values of one-body operators. 
The final scattering angles 
are found by matching the outgoing channel after separation to a pure Coulomb
trajectory~\cite{davies1981}.

Due to numerical complexity and demand of extensive computer time
the early TDHF calculations of deep-inelastic collisions employed approximations and assumptions which were not present in the basic theory, such
as the limitation to an approximate two dimensional collision
geometry and the use of the rotating frame approximation, less accurate lattice
discretization techniques, and less accurate energy density functionals without the spin-orbit interaction~\cite{davies1981,davies1981b}.
Approximations of this type limit the number of degrees of
freedom accessible during a collision and hence the nature and degree of dissipation.
Subsequently, the relaxation of these approximations~\cite{umar1986a,umar1989} have been shown to remedy many of the earlier shortcomings~\cite{lee1987}.
In addition, one-body energy
dissipation extracted from TDHF for low-energy fusion reactions was found to be in agreement with the
friction coefficients based on the linear response theory as well as those in models where
the dissipation was specifically adjusted to describe experiments~\cite{washiyama2009a}. All of these
new results suggest that TDHF dynamics provide a good description of 
heavy-ion collisions.
However, in the mean-field approximation since the collective aspects of the collision dynamics are treated semi-classically the fluctuations of the macroscopic variables are severely inhibited.
To remedy this problem one must go beyond TDHF~\cite{tohyama2002a,simenel2011,lacroix2014}.

Recent TDHF studies have been performed to investigate the charge equilibration 
 in deep-inelastic 
collisions~\cite{iwata2010,simenel2011} and  its impact on the interplay between fusion and transfer reactions~\cite{simenel2008,umar2008a,bourgin2016,vophuoc2016,liang2016,godbey2017}.
Recent investigations  have also shown that the one-body dissipation mechanisms included in TDHF 
were the most relevant to describe fully damped reactions such as quasi-fission 
\cite{wakhle2014,oberacker2014,umar2015a,hammerton2015,umar2016,sekizawa2016}.
In this manuscript we study various aspects of deep-inelastic collisions using the TDHF approach.
To examine the effects of $N/Z$ asymmetry we chose the systems $^{78}$Kr+$^{208}$Pb and $^{92}$Kr+$^{208}$Pb. 
(Intense beams of neutron-rich $^{92}$Kr will be available at future RIB facilities). 
The excitation energy of the primary fragments are calculated using the density-constrained TDHF (DC-TDHF) 
approach~\cite{umar2009a,oberacker2010}.
We also utilize the GEMINI code to study the decay of the primary fragments.
In the next section we give an outline of the theoretical methods employed.
This is followed by the results in Sec.~\ref{sec:results}.
We conclude by giving a summary of findings and future prospects in Sec.~\ref{sec:summary}.

\section{Theoretical Outline}

\subsection{Time-dependent Hartree-Fock method}

The time-dependent Hartree-Fock (TDHF)  theory is a mean-field approximation
of the exact time-dependent many-body problem.
Formally, the strong repulsion
between nucleons at short distances
requires a rearrangement of the standard perturbation theory leading
to the suppression of the strong N-N interaction terms and resulting in
an effective two-body interaction.
The equations of motion governing the nuclear system are derived using
the time-dependent variational principle and lead to the replacement
of the original linear quantum mechanics by a set of coupled non-linear
equations.
The resulting mean-field approximation
yields an excellent description of nuclei throughout the periodic table and
has been successful in the description of the inclusive properties of
low energy heavy-ion collisions.
Generally,
heavy-ion collisions at energies of a few MeV per nucleon above the Coulomb
barrier, are either predominantly fusion or predominantly
deep-inelastic reactions. 
In both cases we are in a regime where classical
descriptions of the relative motion are approximately valid. Thus, TDHF
simulations of these collisions with definite impact parameters are
expected to yield quantitatively good agreement with the corresponding
experimental data.
While TDHF provides a good starting point for a
fully microscopic theory of large amplitude collective
motion~\cite{negele1982,simenel2012,nakatsukasa2016},
only in recent years has it become feasible to perform TDHF calculations on a
three-dimensional Cartesian grid without any symmetry restrictions
and with accurate numerical methods~\cite{reinhard1988,umar1991a,kim1997,maruhn2005,nakatsukasa2005,umar2006c,guo2008,maruhn2014}.
In addition, the quality of energy-density functionals has been substantially
improved~\cite{chabanat1998a,kluepfel2009,kortelainen2010}.

Given a many-body Hamiltonian $\hat{H}$, 
the  action $S$ can be constructed as
\begin{equation}
S=\int_{t_1}^{t_2}dt<\Phi(t)|\hat{H}-i\hbar\partial_t|\Phi(t)>\;.
\end{equation}
Here, $\Phi$ denotes the time-dependent correlated many-body
wavefunction, $\Phi({\bf r_1,r_2,\ldots,r_{A}};t)$. 
The variational principle $\delta S=0$ is then equivalent to
the time-dependent Schr\"odinger equation.
In the TDHF approximation the many-body wavefunction is replaced by a single
Slater determinant and this form is preserved at all times.
The determinental form guarantees the antisymmetry required by the Pauli
principle for a system of fermions. In this limit, the
variation of the action yields the most probable time-dependent mean-field path
between points $t_1$ and $t_2$ in the multi-dimensional
space-time phase space:
\begin{equation}
\delta S=0 \rightarrow \Phi_0(t)\;,
\label{variat}
\end{equation}
where  $\Phi_0(t)$ is  a Slater determinant
with the associated single-particle states $\phi_{\lambda}(\vcr,t)$.
The variation in Eq.(\ref{variat}) is performed with respect to
the single-particle states $\phi_{\lambda}$ and $\phi^{*}_{\lambda}$. 
This leads to a set of coupled, nonlinear, self-consistent initial value equations
for the single-particle states
\begin{equation}
h\left( \left\{ \phi_{\mu} \right\} \right) \phi_{\lambda}=i\hbar
\partial_t{\phi_{\lambda}}
\;\;\;\;\;\;\;\;\;\lambda=1,...,N\;,
\label{tdhf0}
\end{equation}
and their Hermitic conjugates, where $N$ is the number of particles.
These are the fully microscopic TDHF equations.
As we see from Eq.(\ref{tdhf0}), each single-particle state evolves in the
mean-field generated by the concerted action of all the other single-particle
states.

In standard TDHF applications to heavy-ion collisions, the initial nuclei are calculated using the static 
Hartree-Fock (HF) theory and the Skyrme
functional~\cite{chabanat1998a}. The resulting Slater
determinants for each nucleus comprise the larger Slater determinant describing the colliding
system during the TDHF evolution.
Nuclei are assumed
to move on a pure Coulomb trajectory until the initial separation between the nuclear centers used
as initial condition in TDHF evolution.
Of course, no assumption is made on the subsequent  trajectory in the TDHF evolution.
Using the Coulomb trajectory we compute the relative kinetic energy at this
separation and the associated translational momenta for each nucleus. The nuclei are then boosted
by multiplying the HF states with
\begin{equation}
\Phi _{j}\rightarrow \exp (\imath\mathbf{k}_{j}\cdot \mathbf{R})\Phi _{j}\;,
\end{equation}
where $\Phi _{j}$ is the HF state for nucleus $j$ and $\mathbf{R}$ is the corresponding
center of mass coordinate
\begin{equation}
\mathbf{R}=\frac{1}{A_{j}}\sum _{i=1}^{A_{j}}\mathbf{r}_{i}\;.
\end{equation}
The Galilean invariance and the conservation of the total energy in the Skyrme TDHF equations are used to check the
convergence of the calculations.

Since TDHF is based on the independent-particle approximation
it can be interpreted as the semi-classical limit of a fully quantal theory thus allowing
a connection to macroscopic coordinates and providing insight about the collision process.
In this sense the TDHF dynamics can  only be used to compute
the semiclassical trajectories of the collective moments of the composite system as a
function of time. Note that the part of the residual interaction which is neglected in TDHF may produce
fluctuations and correlations which affect these trajectories.
Recent beyond TDHF developments have been used to investigate the effects of such fluctuations in heavy-ion
collisions~\cite{simenel2011,washiyama2009b}.
However, the TDHF approach is optimized to the expectation values of one-body operators~\cite{balian1981}
and is then capable to predict these quantities.
This was demonstrated by the recent successes of TDHF in reproducing various reaction mechanisms in heavy-ion collisions.
Moreover, beyond TDHF calculations remain numerically difficult.
We then restrict the present calculations to the TDHF level.

\subsection{DC-TDHF method and excitation energies}
\label{sec:estar}
The excitation energy and, in particular, its repartition between the fragments
also provide important information on the dissipative nature of the reaction mechanisms~\cite{toke1992,klein1997}. 
In TDHF, thermalization is only partial as it only contains 
 one-body dissipation mechanisms such as nucleon evaporation \cite{chomaz1987,avez2013} and damping of collective energy with (nearly) random
collisions of nucleons with the walls of the mean-field.
One-body energy
dissipation extracted from TDHF for low-energy fusion reactions was also found to be in agreement with the
friction coefficients based on the linear response theory as well as those in models where
the dissipation was specifically adjusted to describe experiments~\cite{washiyama2009a}. 

Based on the strategy proposed in~\cite{cusson1985},  
we recently developed an extension to TDHF theory via the use
of a density constraint to calculate fragment excitation energy of {\it each fragment} directly from the
TDHF time evolution~\cite{umar2009a}.
For this purpose, we divide the conserved TDHF energy into
a collective and intrinsic part, and we
assume that the collective part is primarily determined by
the density $\rho(\mathbf{r},t)$ and the current $\mathbf{j}(\mathbf{r},t)$.
Consequently, the excitation energy can be written in the form
\begin{equation}
E^{*}(t)=E_{\mathrm{TDHF}}-E_{\mathrm{coll}}\left(\rho(t),\mathbf{j}(t)\right)\;,
\label{eq:estar}
\end{equation}
where $E_\mathrm{TDHF}$ is the total energy of the dynamical system, which is a conserved quantity,
and $E_\mathrm{coll}$ represents the collective energy of the system. The collective energy
consists of two parts
\begin{equation}
E_{\mathrm{coll}}\left(t\right)= E_{\mathrm{kin}}\left(\rho(t),\mathbf{j}(t)\right)
 + E_{\mathrm{DC}}\left(\rho(t)\right)\;,
\end{equation}
where $E_{\mathrm{kin}}$ represents the kinetic part and is given by
\begin{equation}
E_{\mathrm{kin}}\left(\rho(t),\mathbf{j}(t)\right)=\frac{m}{2}\int\;{\rm d}^{3}r\;\mathbf{j}^2(t)/\rho(t)\;,
\label{eq:ekin}
\end{equation}
which is asymptotically equivalent to the kinetic energy of the
relative motion, $\frac{1}{2}\mu\dot{R}^2$, where $\mu$ is the
reduced mass and $R(t)$ is the ion-ion separation distance.
The energy $E_\mathrm{DC}$ is the density-constrained TDHF energy, the lowest-energy state of all possible
TDHF states with the same density with no excitation~\cite{umar2006b}.
This gives us new information on the repartition of the excitation energy between the heavy and light fragments
which is not available in standard TDHF calculations, except with projection techniques \cite{sekizawa2015}.

\section{Results}
\label{sec:results}

We have used TDHF theory to study the reactions $^{78}$Kr+$^{208}$Pb and $^{92}$Kr+$^{208}$Pb
at the energy $E=8.5$~MeV/$A$.
TDHF calculations were done in a numerical
Cartesian box which is $65$~fm along the collision axis, $50$~fm in the reaction plane perpendicular to
the reaction axis and $30$~fm in the direction perpendicular to the reaction plane.
The two nuclei are placed at an initial separation of $30$~fm.
Calculations used the SLy4d Skyrme functional~\cite{kim1997} without the pairing interaction as 
described in Ref.~\cite{umar2006c}.
Static calculations are done using the damped-relaxation method~\cite{bottcher1989}.
Krypton nuclei used in these calculations are deformed with deformation parameters $\beta_2=0.088$ for $^{78}$Kr and 
$\beta_2=0.178$ for $^{92}$Kr. To account for this deformation dependence we have performed two sets of calculations
for each Kr nucleus, one with the symmetry axis of the nucleus in the direction of the collision axis $(\beta=0\degree)$
and the other with the symmetry axis perpendicular the collision axis $(\beta=90\degree)$.

\subsection{Main scattering features}

\begin{figure}[!htb]
\includegraphics*[width=8.6cm]{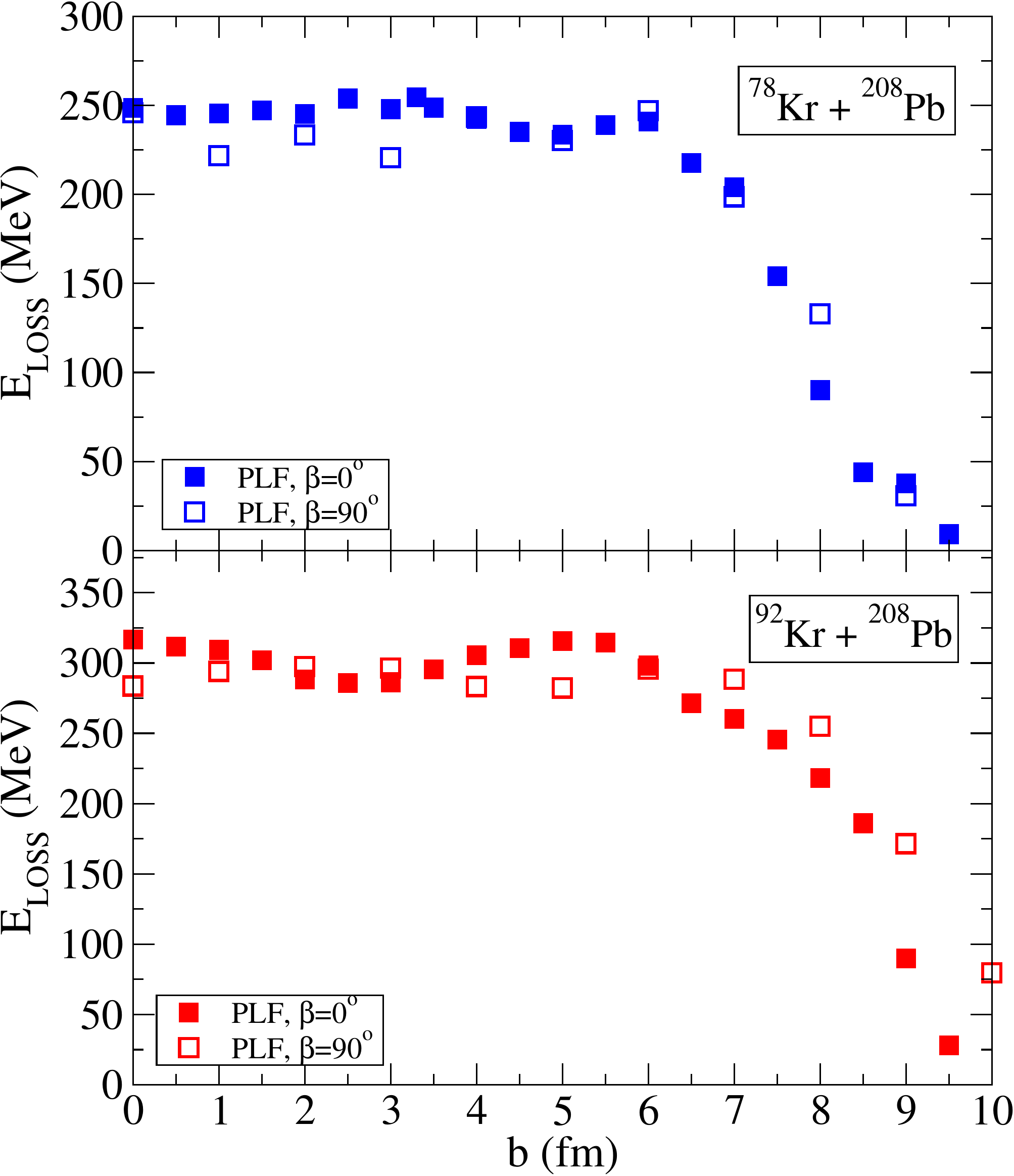}
\caption{\protect (Color online) Energy loss, $E_\mathrm{loss}$ (MeV) versus impact parameter $b$~(fm)
for the reactions $^{78}$Kr+$^{208}$Pb and $^{92}$Kr+$^{208}$Pb
at $E=8.5$~MeV/$A$. Angle $\beta$ represents the initial
orientation of the deformed $^{78}$Kr nucleus with respect to the beam axis.
\label{bvsEloss}}
\end{figure}
\begin{figure}[!htb]
\includegraphics*[width=8.6cm]{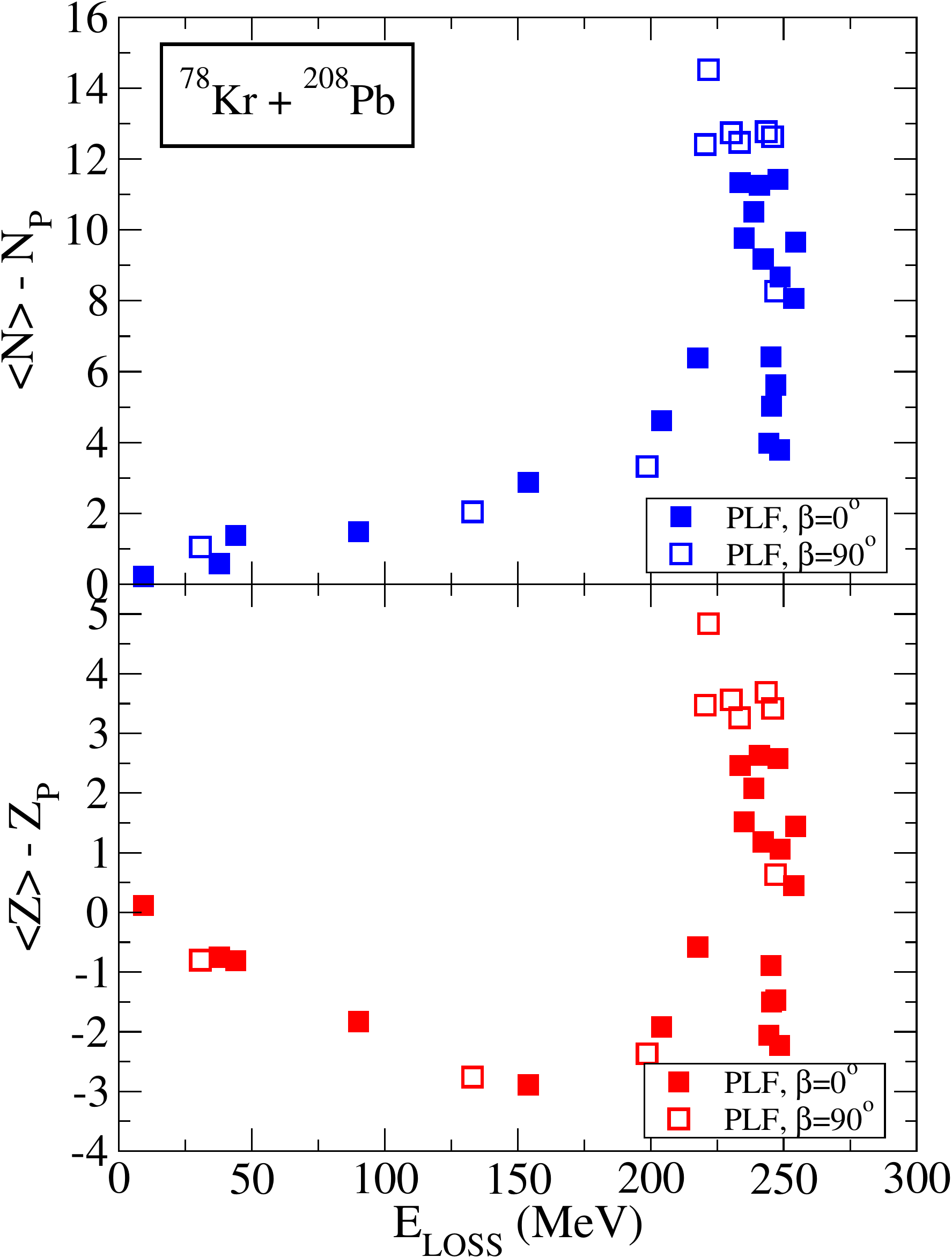}
\caption{\protect (Color online) Neutron and proton numbers transferred to/from the PLF for the reaction $^{78}$Kr+$^{208}$Pb
at $E=8.5$~MeV/$A$ as a function of $E_\mathrm{loss}$. Angle $\beta$ represents the initial
orientation of the deformed $^{78}$Kr nucleus with respect to the beam axis.
\label{78KrNPvsEloss}}
\end{figure}
\begin{figure}[!htb]
\includegraphics*[width=8.6cm]{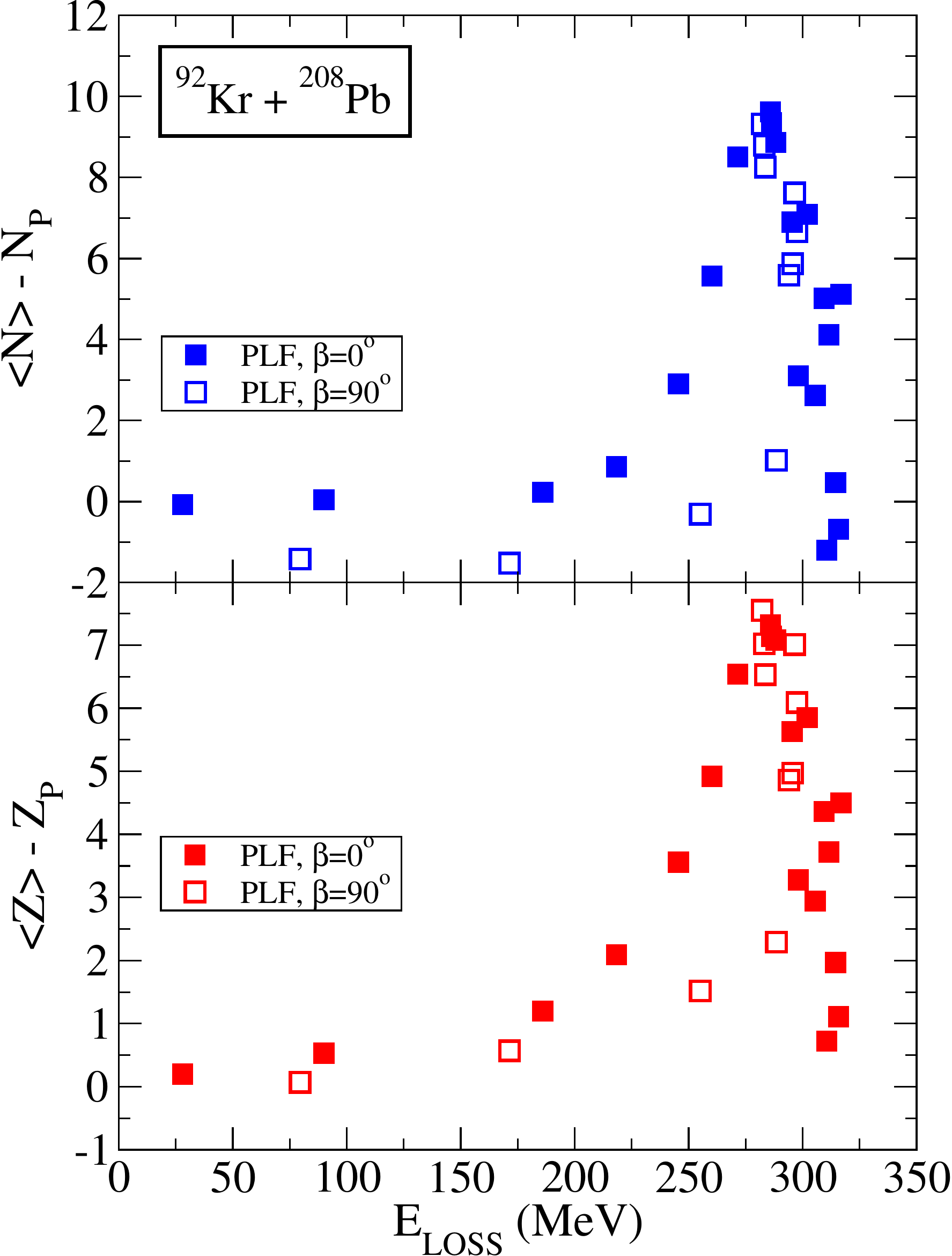}
\caption{\protect (Color online) Neutron and proton numbers transferred to/from the PLF for the reaction 
        $^{92}$Kr+$^{208}$Pb
at $E=8.5$~MeV/$A$ as a function of $E_\mathrm{loss}$. Angle $\beta$ represents the initial
orientation of the deformed $^{92}$Kr nucleus with respect to the beam axis.}
\label{92KrNPvsEloss}
\end{figure}

In Fig.~\ref{bvsEloss} we plot the kinetic energy loss, $E_\mathrm{loss}$~(MeV), versus impact parameter, $b$~(fm),
for both  reactions. 
Energy loss is defined as the difference in initial c.m. energy and final c.m. energy of the outgoing fragments.
Angle $\beta$ represents the initial
orientation of the deformed Kr nucleus with respect to the beam axis as discussed above.
We note that for both systems there is a plateau for the energy loss for impact parameters
up to about $b=6$~fm. For larger impact parameters energy loss gradually decreases as expected.
We also note that the energy loss for the neutron-rich $^{92}$K reaction is considerably larger
than the one for the $^{78}$Kr collision.
This can be interpreted as an effect of the higher  beam energy in the  $^{92}$Kr+$^{208}$Pb reaction. 
(Both reactions have the same beam energy per nucleon of 8.5~MeV/$A$, but $^{92}$Kr is $\sim 15\%$ heavier and then more energetic than $^{78}$Kr.)
Another interesting point is that the initial orientation of the Kr nuclei seem to have minimal effect
on the energy loss for the reaction.
Figures~\ref{78KrNPvsEloss} and \ref{92KrNPvsEloss} show the number of neutrons and protons transferred to/from the
projectile like fragment (PLF) for the reactions $^{78}$Kr+$^{208}$Pb and $^{92}$Kr+$^{208}$Pb,
respectively. As anticipated for small energy losses originating from larger impact parameters
the transfer of nucleons diminishes. However, for larger energy loses we notice important
differences between the two systems. In the case of $^{78}$Kr+$^{208}$Pb system, neutron transfer
to the PLF gradually increases with increasing energy loss and peak around $E_\mathrm{loss}\approx 250$~MeV.
At this energy loss, corresponding to the plateau region of Fig.~\ref{bvsEloss} we see a large range of
$4-15$ neutrons transferred to the PLF. An important point to notice is that the transfer is unidirectional,
namely to the PLF only. The orientation effects of the $^{78}$Kr is more pronounced with the $(\beta=90\degree)$
orientation resulting in larger transfers at the maximum energy loss.
The situation for proton transfer is more complicated than that for neutrons. For small energy losses the
protons seem to be transferred from the $^{78}$Kr to the $^{208}$Pb for up to about $E_\mathrm{loss}\approx 210$~MeV,
subsequently changing direction. At the maximum energy loss region the transfers for the $(\beta=90\degree)$
degree orientation is from $^{208}$Pb to $^{78}$Kr, whereas the $\beta=0\degree$ orientation results in
transfers in both directions. One may understand this behavior in terms of $N/Z$ equilibration as follows;
for small energy losses and small transfers the neutron poor $^{78}$Kr nucleus can equilibrate faster by
giving away some protons and receiving a small number of neutrons. However, as it receives more and more
neutrons it no longer has to give out protons and actually now it can receive some protons. A more detailed
discussion about the behavior in the maximum energy loss region will be given below.
Figure~\ref{92KrNPvsEloss} shows the neutron and proton transfers to/from the PLF for the $^{92}$Kr+$^{208}$Pb
system. In this case most of the neutron and all of the proton transfer is from $^{208}$Pb to $^{92}$Kr.
Only for the $(\beta=90\degree)$ orientation of $^{92}$Kr we see a region of energy loss where the a few neutrons
are transferred in the opposite direction.
In the region of small energy loss we see no appreciable neutron transfer to the PLF while a small proton
transfer takes place. For larger energy losses the number of transferred protons increases followed by the
transfer of neutrons. Around the maximum energy loss, $E_\mathrm{loss}\approx 300$~MeV, a wide distribution
of transfers occur. Again, this behavior can largely be explained as a dynamical $N/Z$ equilibration. 

\begin{figure}[!htb]
\includegraphics*[width=8.6cm]{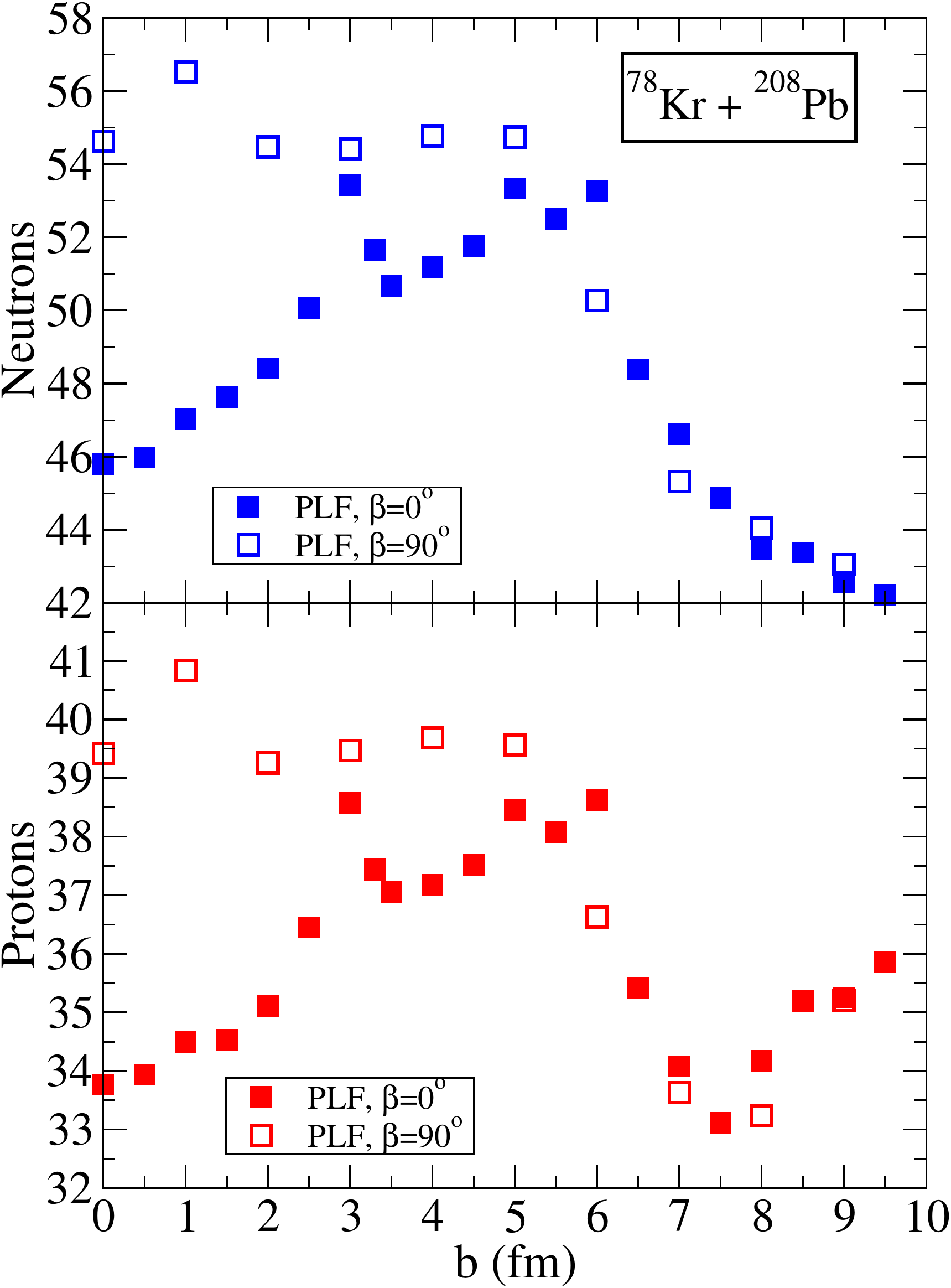}
\caption{\protect (Color online) Neutron and proton numbers of the PLF for the reaction $^{78}$Kr+$^{208}$Pb
at $E=8.5$~MeV/$A$ as a function of impact parameter $b$~(fm). Angle $\beta$ represents the initial
orientation of the deformed $^{78}$Kr nucleus with respect to the beam axis.
\label{78KrNPvsb}}
\end{figure}

\begin{figure}[!htb]
\includegraphics*[width=8.6cm]{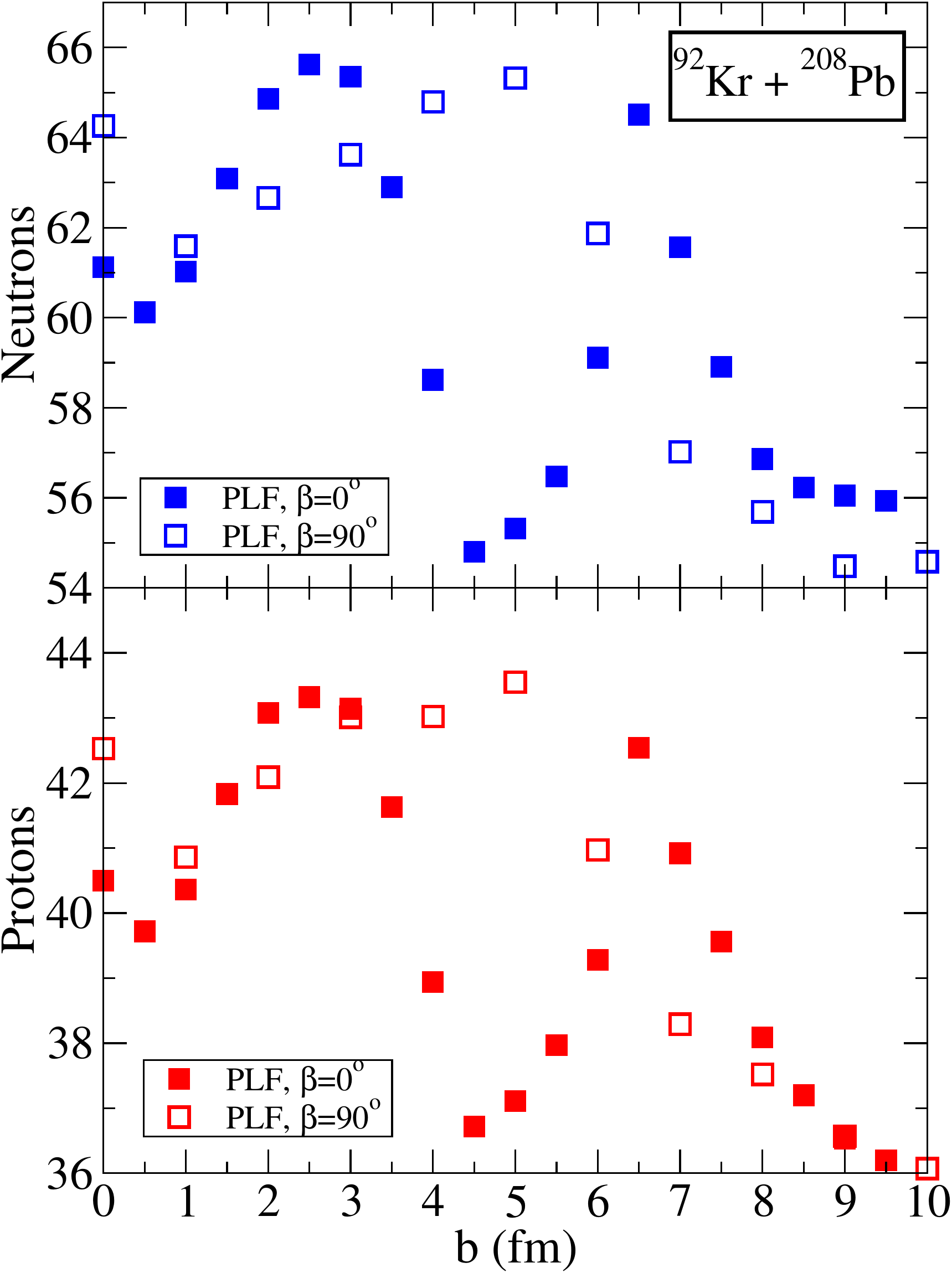}
\caption{\protect (Color online) Neutron and proton numbers of the PLF for the reaction $^{92}$Kr+$^{208}$Pb
at $E=8.5$~MeV/$A$ as a function of impact parameter $b$~(fm). Angle $\beta$ represents the initial
orientation of the deformed $^{92}$Kr nucleus with respect to the beam axis.}
\label{92KrNPvsb}
\end{figure}

In order to
gain more insight about the dependence of transfer on reaction dynamics we can investigate the impact parameter dependence
of these reactions. In Fig.~\ref{78KrNPvsb} we plot the neutron and proton numbers of the PLF for the 
reaction $^{78}$Kr+$^{208}$Pb
at $E=8.5$~MeV/$A$ as a function of impact parameter.
Here, the dependence of transfer on the Kr orientation angle $\beta$ is much more pronounced.
While for larger impact parameters ($b>6$~fm) the dependence on orientation is negligible,
for smaller impact parameters we observe much larger neutron and proton transfer for the $\beta=90\degree$
orientation of $^{78}$Kr. 
As a matter of fact for central impact parameters ($b<3$~fm)
the transfer of protons to PLF occur in opposite direction for the two orientations. For these impact
parameters $\beta=90\degree$ orientation has a large neutron and proton transfers to the PLF, whereas
the $\beta=0\degree$ orientation actually looses protons and gains a few neutrons.
For large impact parameters ($b>6$~fm) there is proton transfer from $^{78}$Kr, which reaches maximum
at $b=7.5$~fm.
The corresponding plot of neutron and proton numbers of the PLF for the $^{92}$Kr+$^{208}$Pb system is 
shown in Fig.~\ref{92KrNPvsb}.
For this system, for central impact parameters, we see an increase in transfer as the impact parameter increases
for both orientations. Subsequently, for $\beta=0\degree$ orientation there is a drop in both the neutron
and proton curves around $b=4.5$~fm, practically going down to no transfer. In contrast transfer for 
$\beta=90\degree$ orientation remains high in this region. For larger impact parameters transfer for both
orientations decrease gradually.
One can characterize the structures seen in the impact parameter dependence as being comprised of
fine-structures and gross-structures. These structures emanate from a complicated amalgamation of
microscopic shell-structure and collective dynamics and show that such dependencies are not always
amenable to phenomenological modeling. 
For example, for certain combinations of angular momentum and energy, two cluster
orbits in the separating fragments may have a large overlap in
momentum space, which substantially enhances the probability for
the transfer of particles.
Furthermore, they may not necessarily 
be indicated in experimentally observed quantities.

\begin{figure}[!htb]
	\includegraphics*[width=8.6cm]{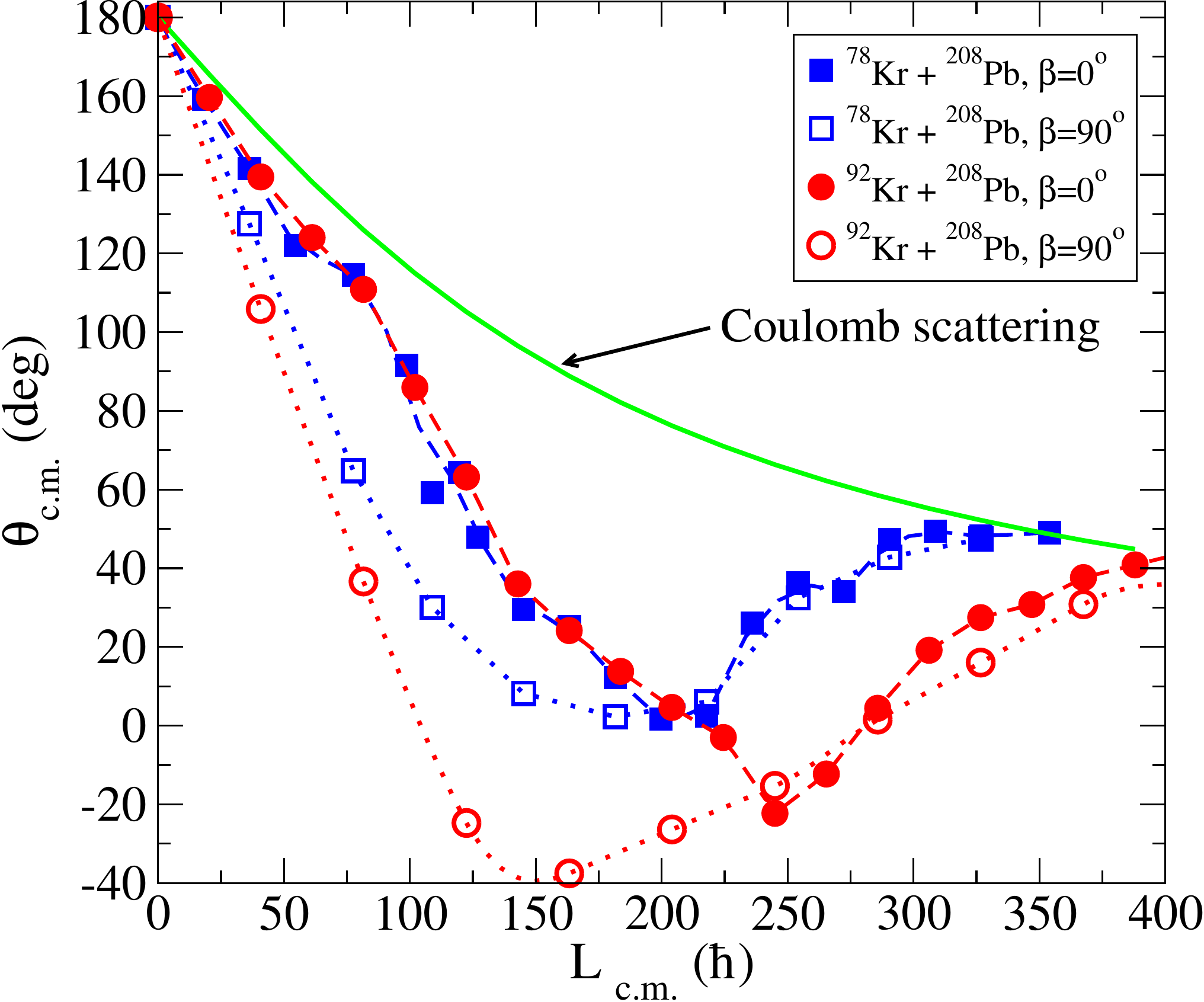}
	\caption{\protect (Color online) The deflection function for the reactions $^{78}$Kr+$^{208}$Pb 
		and $^{92}$Kr+$^{208}$Pb
		at $E=8.5$~MeV/$A$. Angle $\beta$ represents the initial
		orientation of the deformed Kr nuclei with respect to the beam axis.
		As a reference we also show the Rutherford scattering deflection angle (green curve).}
	\label{Theta_vs_L}
\end{figure}

Dissipative aspects of a deep-inelastic reaction are often studied in terms of the
dependence of energy loss on the deflection angle.
The deflection function (the c.m. scattering angle versus the initial orbital angular
momentum $L_\mathrm{c.m.}$) is related to the differential cross-section and the dependence of the final
kinetic energy of the fragments, versus the scattering angle.
The contour plot of constant cross section in the deflection angle and kinetic energy plane is called Wilczynski plot~\cite{wilczynski1973}.
In Fig.~\ref{Theta_vs_L} we show the deflection function
plotted as a function of the initial orbital angular momentum.
For reference we also show the pure Rutherford scattering deflection angle (green curve).
As we see for head-on collisions ($L_\mathrm{c.m.}=0$) and for the most peripheral collisions the
deflection function approaches the Rutherford scattering limit.
In the intermediate region of partial waves the balance of the attractive nuclear
force and repulsive Coulomb and centrifugal forces determine the behavior.
A stronger deflection due to nuclear orbiting is observed in the collisions induced by the $^{92}$Kr beam. 
This can be interpreted as an effect of larger angular momentum in these collisions. 
Indeed, the partial wave corresponding to the grazing angle can be
obtained using the sharp-cutoff model to be $518\hbar$ for $^{78}$Kr
and $596\hbar$ for $^{92}$Kr.

\begin{figure}[!htb]
	\includegraphics*[width=8.6cm]{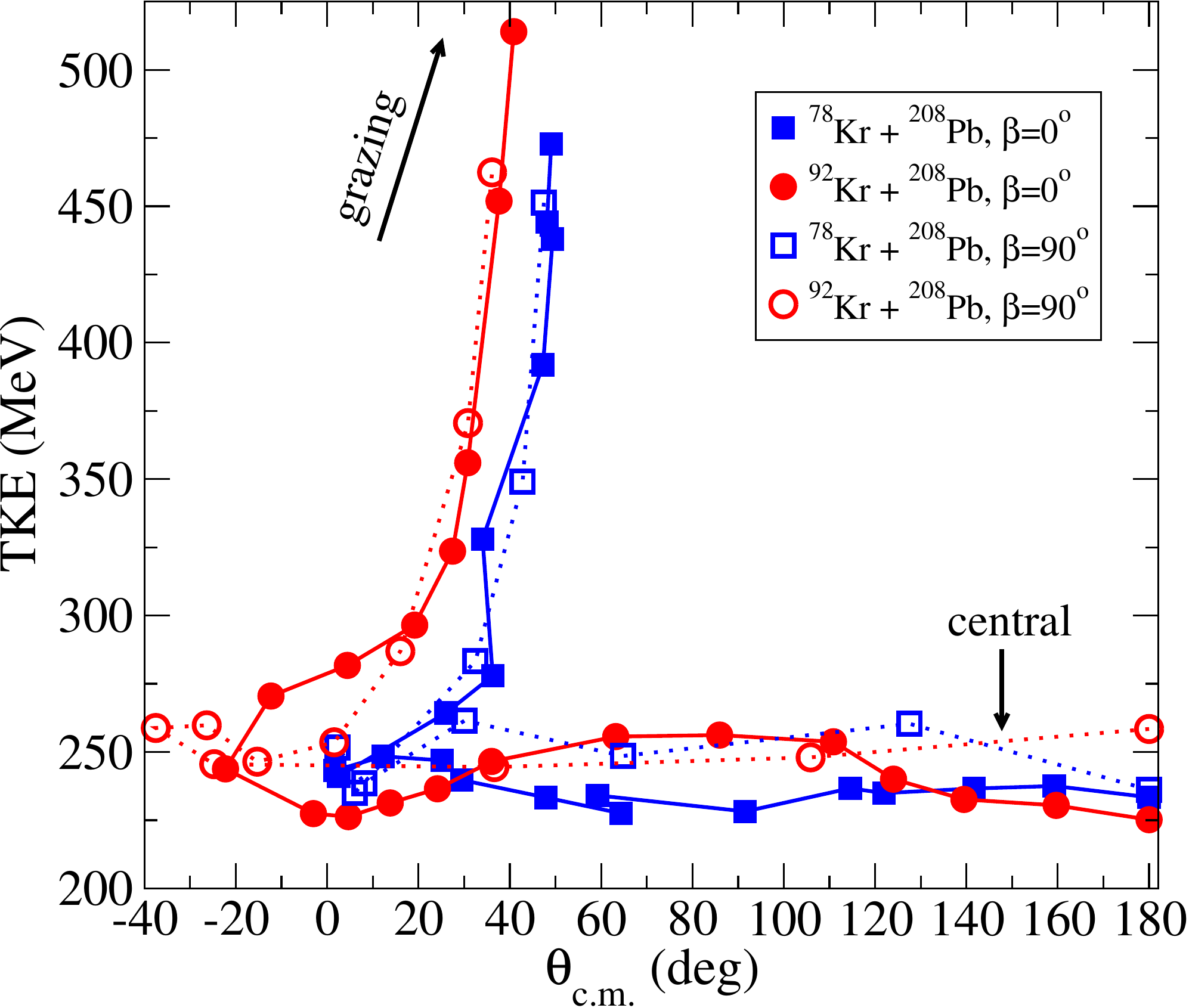}
	\caption{\protect (Color online) The final kinetic energy versus the scattering angle for the reactions 
    $^{78}$Kr+$^{208}$Pb and $^{92}$Kr+$^{208}$Pb
    at $E=8.5$~MeV/$A$. Angle $\beta$ represents the initial
	orientation of the deformed Kr nuclei with respect to the beam axis.}
	\label{TKE_vs_theta}
\end{figure}

Figure~\ref{TKE_vs_theta} shows the TKE versus scattering angle.
Fully damped collisions produce fragments with the same TKE in both reactions, 
with a strong orbiting (spanning all angles) characteristic of deep-inelastic collisions. 
Figures~\ref{Theta_vs_L} and \ref{TKE_vs_theta} demonstrate the strong correlation of
the energy loss with the deflection angle and accordingly with the impact parameter.
These results show that TDHF calculations reproduce many of the main features of deep-inelastic
collision phenomena.
In the language of models these are related to the phenomena of orbiting incorporating
friction.
Finally, we should mention that the TDHF results yield single curves which should be compared with the most
probable experimental energy-angle correlation, the maximum of the contours.

\subsection{N/Z equilibration}

The influence of isospin flow during strongly-damped heavy-ion reactions is usually discussed in term of the
$(N/Z)$ asymmetry of the target and projectile. In the current study the $(N/Z)$ values for $^{78}$Kr
and  $^{92}$Kr are 1.16 and 1.55, respectively. The $(N/Z)$ value for $^{208}$Pb is 1.54.
This implies that the $^{92}$Kr+$^{208}$Pb reaction is a nearly $(N/Z)$ symmetric collision.
The $(N/Z)$ values for the compound systems for $^{78}$Kr+$^{208}$Pb and $^{92}$Kr+$^{208}$Pb
are 1.42 and 1.54, respectively. For a $(N/Z)$ asymmetric system the fragments emerging from a
deep-inelastic collision should have their average somewhere between the $(N/Z)$ values of the
target and the projectile, depending on the degree of equilibration.
Naturally, the amount of equilibration depends on the energy and impact parameter, which determine
the amount of time the system spends in a di-nuclear configuration.
\begin{figure}[!htb]
    \includegraphics*[width=8.6cm]{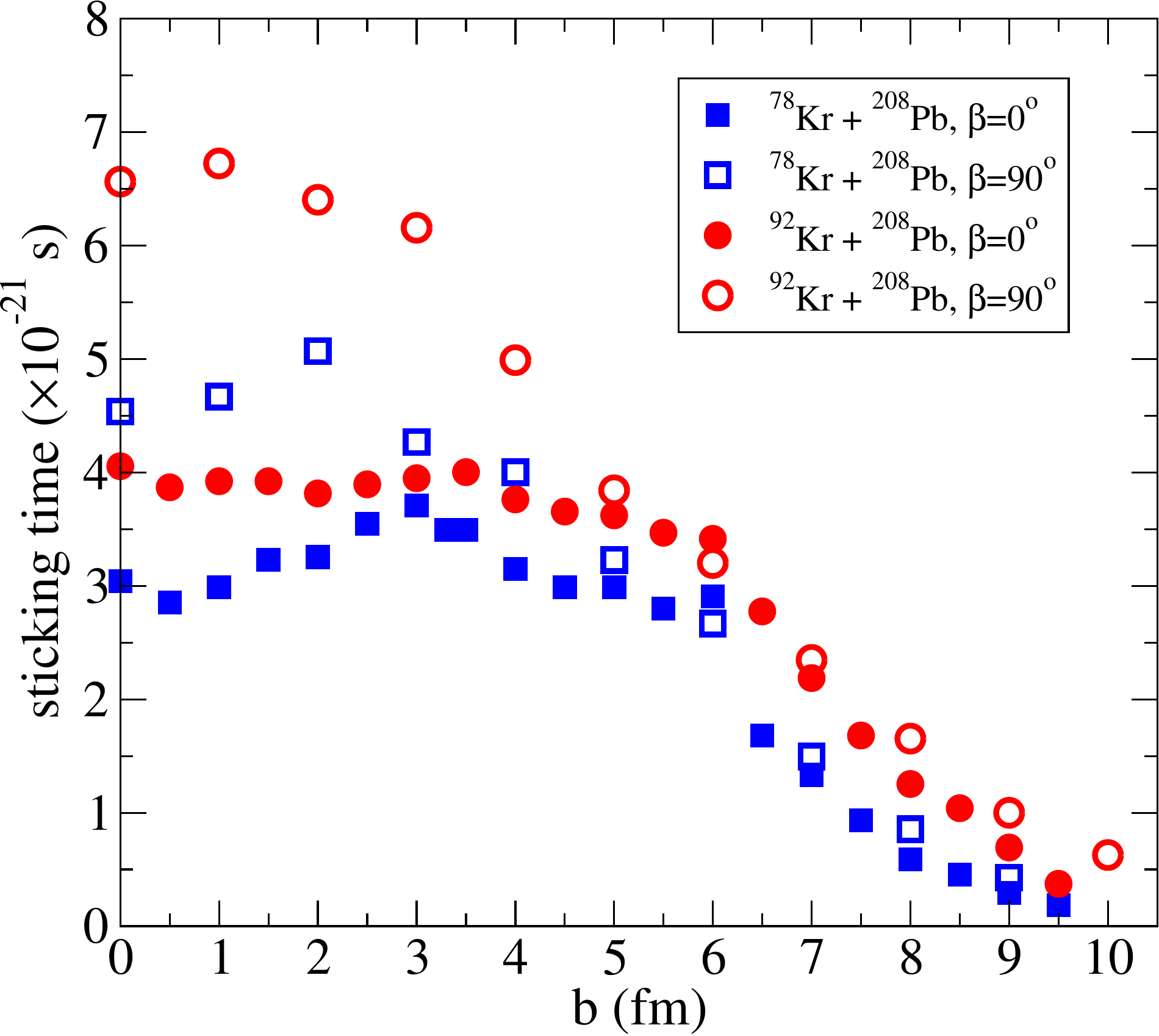}
    \caption{\protect (Color online) The sticking time as a function of impact parameter for 
        $^{78}$Kr+$^{208}$Pb and $^{92}$Kr+$^{208}$Pb
        at $E=8.5$~MeV/$A$. Angle $\beta$ represents the initial
        orientation of the deformed Kr nuclei with respect to the beam axis.}
    \label{sticking_t}
\end{figure}
In Fig.~\ref{sticking_t} we plot the sticking time (time spent from the initial contact to final separation) as a 
function of impact parameter and initial orientation of Kr.
First, we observe the obvious, namely the most peripheral collisions are fast and have the least time
for equilibration, whereas the central impact parameters allow for much longer equilibration times.
We also observe that for central collisions the dependence of sticking time on the orientation of the Kr nuclei is
evident, with perpendicular orientation resulting in much longer sticking times.
It is very interesting to compare these sticking times with the final $(N,Z)$ content of the fragments 
depicted in Figs.~\ref{78KrNPvsb} and \ref{92KrNPvsb}. For example, the structures observed in $^{92}$Kr+$^{208}$Pb
collision for $\beta=0\degree$ orientation around $b=4.5$~fm region cannot simply be explained by the
sticking time which is relatively smooth in this region. This suggests that for such collisions while
the sticking time plays a certain role in determining the reaction products shell effects are still
very important. This is one of the reasons why modeling of these reactions based on general macroscopic assumptions
may not always be appropriate.
\begin{figure}[!htb]
	\includegraphics*[width=8.6cm]{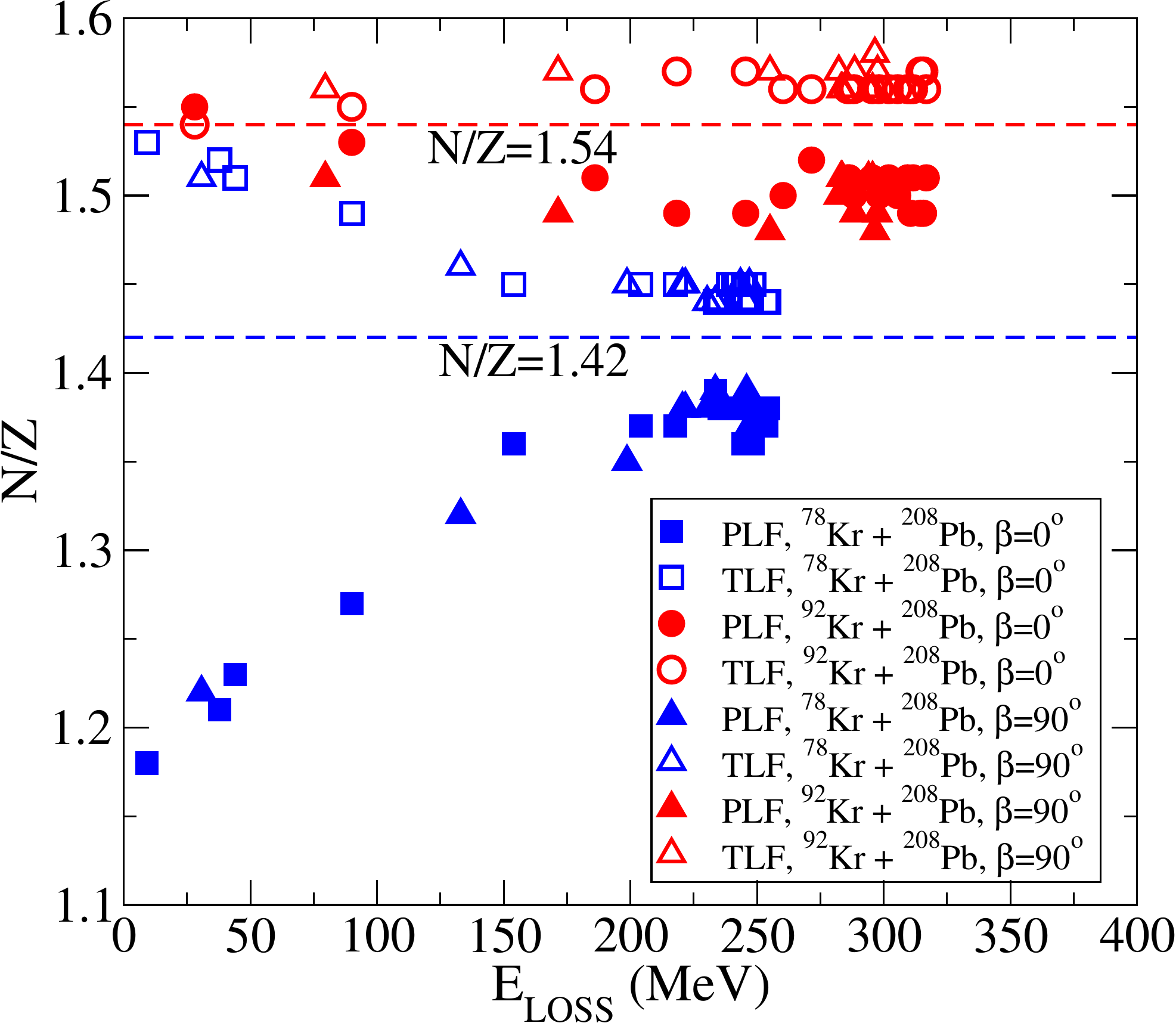}
	\caption{\protect (Color online) The $(N/Z)$ values for PLF and TLF as a function of 
		$E_\mathrm{loss}$ for $^{78}$Kr+$^{208}$Pb and $^{92}$Kr+$^{208}$Pb
		at $E=8.5$~MeV/$A$. Angle $\beta$ represents the initial
		orientation of the deformed Kr nuclei with respect to the beam axis.
	    The horizontal lines represent the $(N/Z)$ values for the compound system for 
        the two reactions.}
	\label{NoZ}
\end{figure}

The measure of $(N/Z)$ equilibration during the reaction is shown in Fig.~\ref{NoZ} in
terms of the $(N/Z)$ values of the PLF and target-like fragment (TLF).
As expected, we observe that for 
peripheral collisions, the
PLF and TLF $(N/Z)$ values are close to the projectile/target values, respectively.
For the $^{78}$Kr+$^{208}$Pb reaction, these values are quite different, and thus a large degree of charge equilibration is observed in more damped collisions, producing fragments with $(N/Z)$ values close (but not equal to) the $(N/Z)$ of the compound system.
Indeed, the PLF and TLF $(N/Z)$ values in $^{78}$Kr+$^{208}$Pb fully damped collisions are $\sim 1.38$ an $\sim 1.44$, respectively, while the compound system has $N/Z\simeq 1.42$.

In contrast, for the $^{92}$Kr+$^{208}$Pb system, we observe only small deviations around the initial $(N/Z)$ values of the fragments as the system is already close to equilibrium as far as the isospin degree of freedom is concerned. 
Looking into details, it is interesting to note that for strongly damped collisions, the $(N/Z)$ values in the fragments become slightly more asymmetric than in the entrance channel ($\sim 1.56$ for the PLF and $\sim 1.50$ for the TLF). 

The fact that $(N/Z)$ values of the fragments are not exactly equal, even in fully damped collisions, 
is not a signature for being out of isospin equilibrium~\cite{jedele2017}. 
Indeed, the thermodynamic equation of state indicates that the fragments should approach a common chemical potential 
(in our case strongly affected by symmetry energy) rather than a common composition. 
In fact, the $(N/Z)$ value provides only an approximate proxy for the chemical potential which depends on variations in internal energy, density and ground-state binding energies. 

\begin{figure}[!htb]
	\includegraphics*[width=8.6cm]{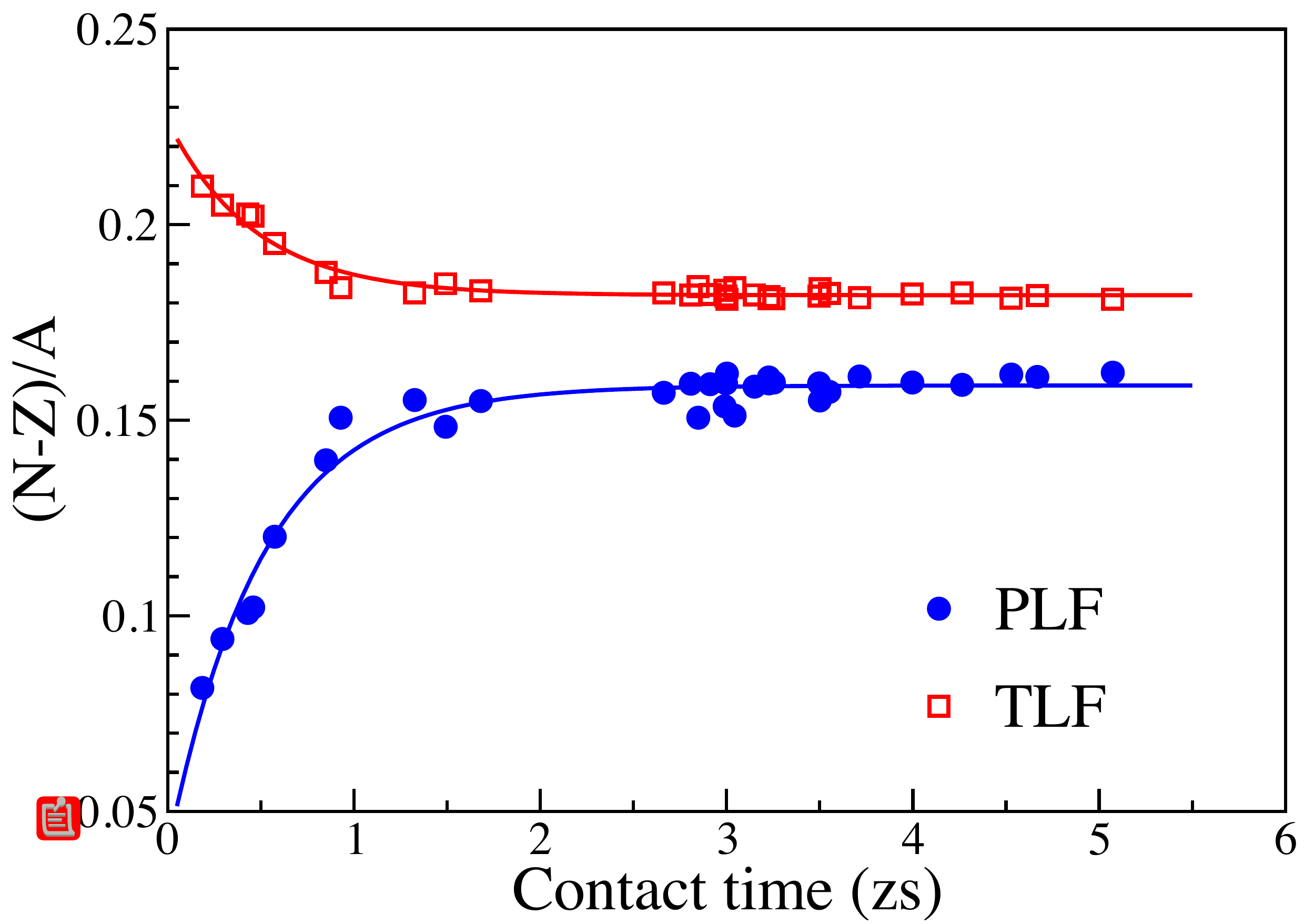}
	\caption{\protect (Color online) The  $(N-Z)/A$ value of the primary PLF (full circles) and TLF (open squares)
	formed in $^{78}$Kr+$^{208}$Pb 
    at $E=8.5$~MeV/$A$ are plotted as a function of the contact time 
    between the collision partners. 
    The solid lines show fits to the TDHF results (see text). }
	\label{fig:NZeq}
\end{figure}

Nevertheless, the neutron and proton composition dependence with the contact time can be use to estimate the charge equilibration time.
Charge equilibration is often achieved within about 1~zs, as shown by earlier TDHF calculations~\cite{simenel2012b}.
Figure~\ref{fig:NZeq} shows the evolution of $(N-Z)/A$ as a function of contact time $T$. 
An equilibration time $\tau\sim 0.5$~zs is obtained from the fit $(N-Z)/A=\alpha+\beta\,\exp(-T/\tau)$.
Recently, Jedele et al. \cite{jedele2017} obtained a slightly faster equilibration time 
of $\sim 0.3$~zs from experimental data at Fermi energy. 
The fact that TDHF gives a charge equilibration time of the same order indicates that it incorporates  the essential physics 
to describe this process. 
This also indicates that the charge equilibration mechanisms ought to be similar at different energy regimes. 


\subsection{From primary fragments to cold residues}

If we neglect the fast nucleon evaporation occurring before the last time iteration of the TDHF calculation, 
the calculated PLF and TLF fragments  by TDHF correspond to  primary fragments. 
The absence of quantal decay (other than evaporation of single-particle wave-functions) 
and transitions prevents us from dynamically
calculating the secondary and further fragments. It is expected that the primary fragments
will undergo various processes to approach the $\beta-$stability line in time~\cite{planeta1990}.
However, assuming that it is statistical in character, the deexcitation process
can be calculated with programs like GEMINI~\cite{charity2008,charity2010} provided realistic 
inputs can be obtained.
In addition to the mass and charge of the PLF we calculate the excitation energy of each primary fragment
(discussed in the next section) as well as the angular momentum ``loss'' 
(i.e., transfer from initial orbital angular momentum to intrinsic angular momentum of the fragments). Assuming a division of
remaining angular momentum in proportion to PLF mass one has all the ingredients to employ
the GEMINI code to calculate the deexcitation process (other parameters set to default values).
The methods described in Refs.~\cite{planeta1990,breuer1983} were applied to obtain the centroids 
for the $Z$ and $N$
distributions of post-evaporative projectile-like fragments.

\begin{figure}[!htb]
	\includegraphics*[width=8.6cm]{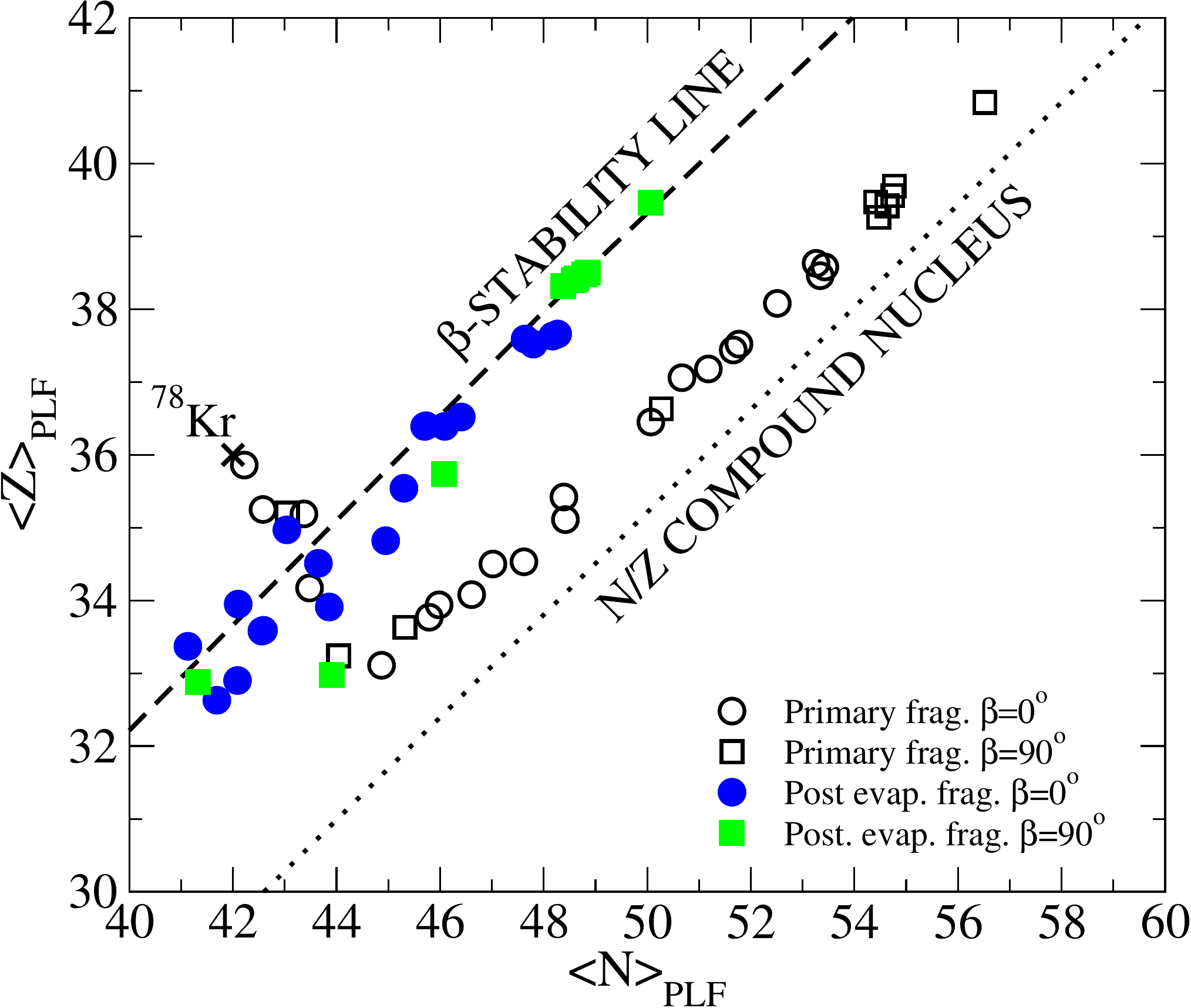}
	\caption{\protect (Color online) The distribution of PLF  neutron and proton numbers plotted in
		the $N-Z$ plane for the $^{78}$Kr+$^{208}$Pb system at $E=8.5$~MeV/$A$. Angle
	    $\beta$ represents the initial orientation of the deformed Kr nuclei with respect to the beam axis.
	    The initial $^{78}$Kr position is marked with $\times$ . The unfilled symbols indicate the
	    primary fragments, while the filled symbols show the fragments after deexcitation.}
	\label{78Kr_NZ_final}
\end{figure}
Similar to previous studies~\cite{planeta1990} in Fig.~\ref{78Kr_NZ_final} we show the evolution of 
the centroids of the
nuclides distributions in the $N$ versus $Z$ plane for different energy-loss bins for
the $^{78}$Kr+$^{208}$Pb system.
The initial $^{78}$Kr position is marked with~$\times$ . The unfilled symbols indicate the
primary fragments, while the filled symbols show the fragments after deexcitation.
We also show the line corresponding to the compound nucleus $(N/Z)$ value of 1.42,
and the $\beta-$stability line.
As we have also observed in Fig.~\ref{NoZ} the primary fragments corresponding to strongly damped
collisions approach $(N/Z)$ values close to the compound nucleus line.
Also shown in Fig.~\ref{78Kr_NZ_final} are the deexcited fragments (filled circles) calculated
with the GEMINI deexcitation code. We see that the deexcited fragments congregate on and around
the $\beta-$stability line. As expected the primary fragments with higher excitation originating
from the strongly damped collisions have a better chance for deexciting to the $\beta-$stability line.
\begin{figure}[!htb]
	\includegraphics*[width=8.6cm]{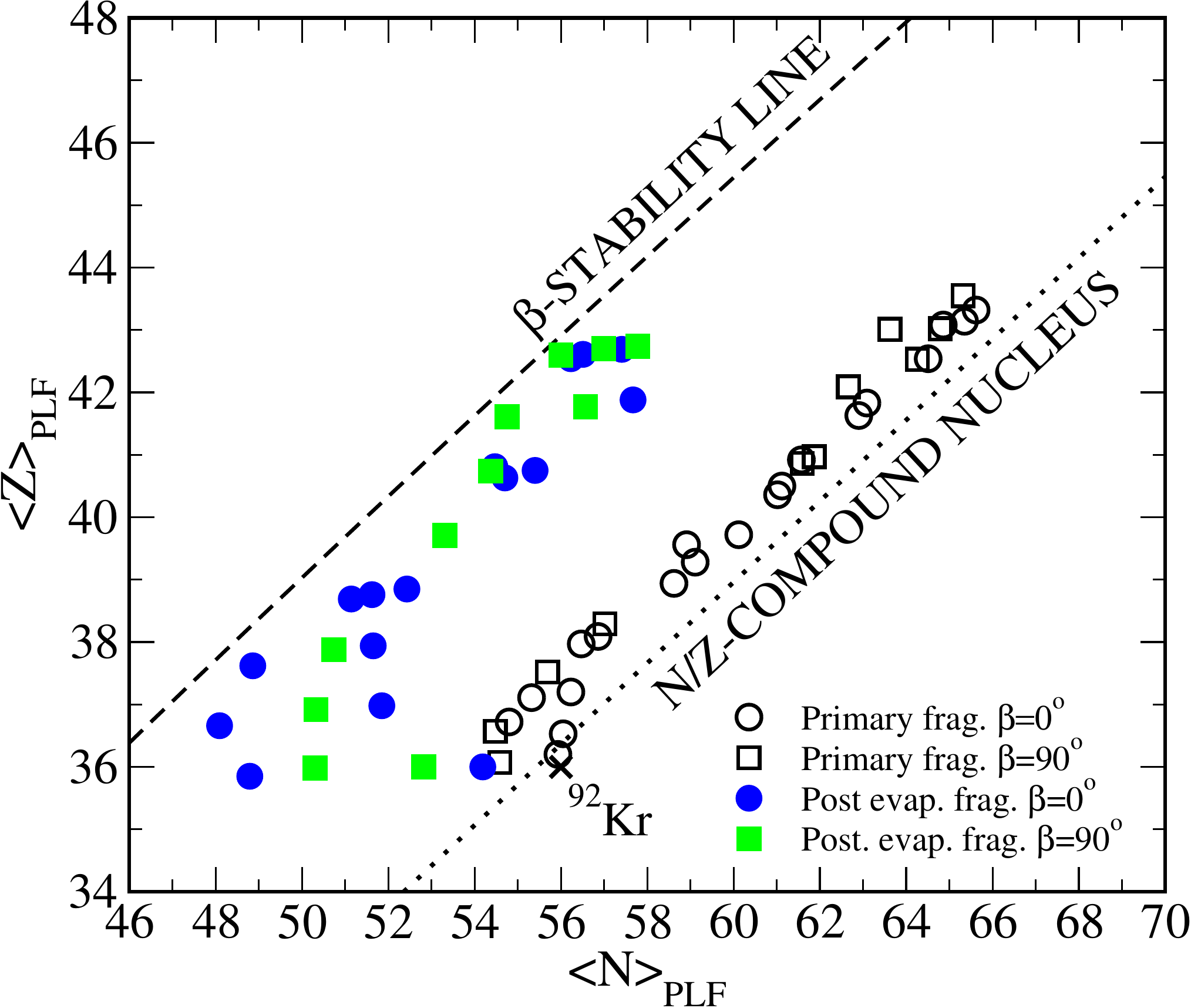}
	\caption{\protect (Color online) The distribution of PLF  neutron and proton numbers plotted in
		the $N-Z$ plane for the $^{92}$Kr+$^{208}$Pb system at $E=8.5$~MeV/$A$. Angle
		$\beta$ represents the initial orientation of the deformed Kr nuclei with respect to the beam axis.
		The initial $^{92}$Kr position is marked with $\times$ . The unfilled symbols indicate the
		primary fragments, while the filled symbols show the fragments after deexcitation.}
	\label{92Kr_NZ_final}
\end{figure}
In Fig.~\ref{92Kr_NZ_final} we show the evolution of the centroids of the
nuclides distributions in the $N$ verse $Z$ plane for different energy-loss bins for
the $^{92}$Kr+$^{208}$Pb system.
The initial $^{92}$Kr position is marked with symbol~$\times$ . Again, the unfilled symbols indicate the
primary fragments, while the filled symbols show the fragments after deexcitation.
The $\beta-$stability line and the line corresponding to the compound nucleus $(N/Z)$ value of 1.54 are
also shown.
Compared to $^{78}$Kr+$^{208}$Pb system the primary fragments produced in $^{92}$Kr+$^{208}$Pb are more
neutron rich, meaning that they are further away from the $\beta-$stability line. Also, one can notice a similar
decay chain length for the excited primary fragments in the two reactions. As a result, the post-evaporative
fragments in $^{92}$Kr+$^{208}$Pb are more neutron rich; that is, they are slightly further from the
$\beta-$stability line in comparison with the $^{78}$Kr+$^{208}$Pb system.

\subsection{Excitation energies}

One of the most interesting aspects of strongly damped collisions is the partial transformation of the
initial available energy into various forms of excitation via dissipative (heat) or non-dissipative
processes, such as deformation and spin of the fragments~\cite{toke1992}.
The excitation energy division between the PLF and TLF is intimately related to ($N/Z)$ equilibration,
degree to which thermal equilibrium is reached, and relaxation times.
In this section, we discuss the excitation properties of the produced primary fragments
using the method described in Sec.~\ref{sec:estar}.

\begin{figure}[!htb]
	\includegraphics*[width=8.6cm]{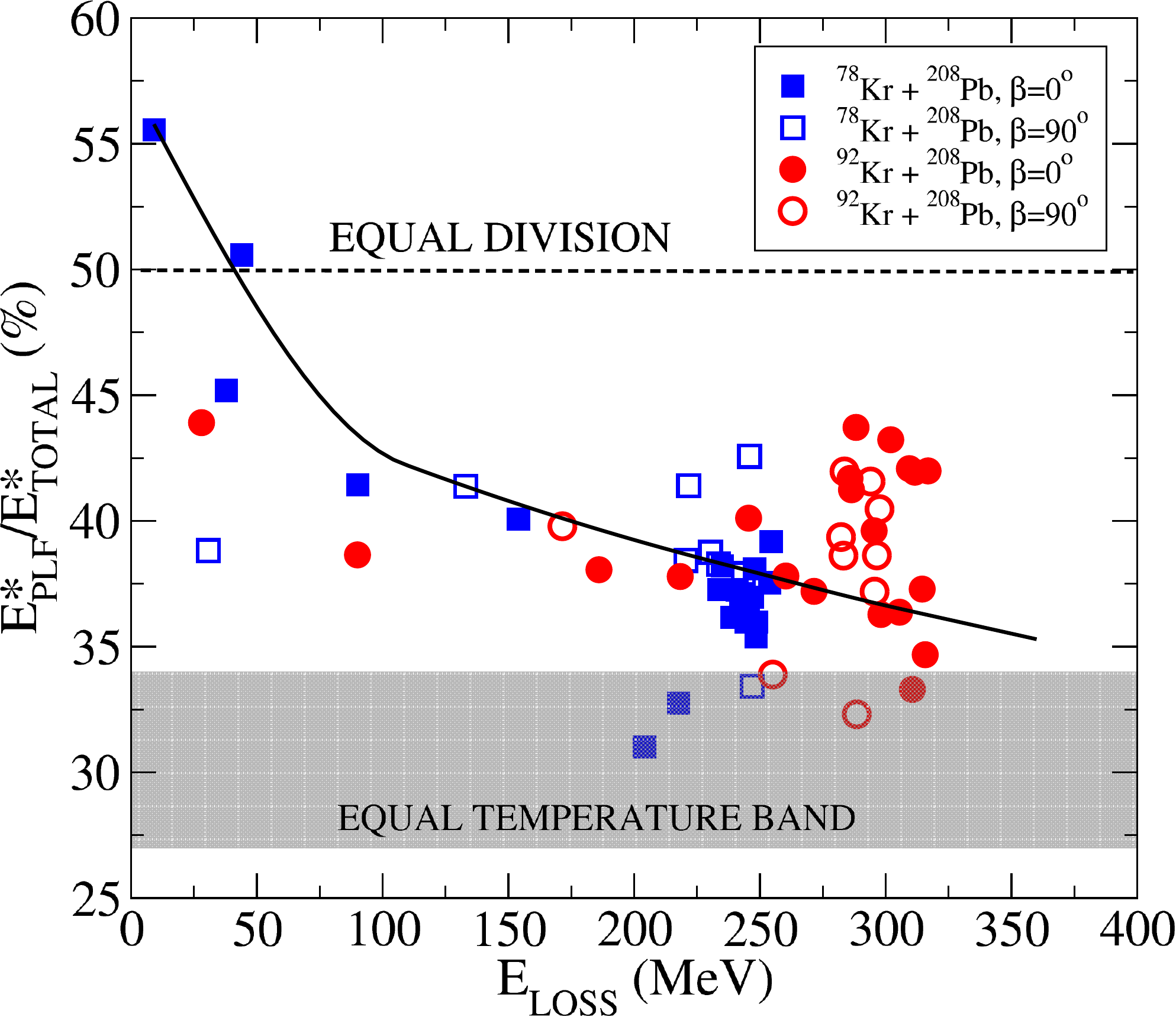}
	\caption{\protect (Color online) 
		Percent of total excitation energy carried by the PLF as a function of $E_\mathrm{loss}$. Angle
		$\beta$ represents the initial orientation of the deformed Kr nuclei with respect to the beam axis.
		The dashed horizontal line marks equal sharing of excitation energy.
        The lower banded region indicates the full thermalization limit.\label{EsPLFoEsT}}
\end{figure}

In Fig.~\ref{EsPLFoEsT} we show the percent fraction of the excitation energy carried by the PLF
as a function of $E_{loss}$ for all the systems studied.
The solid line is drawn by hand to show the general trend of the results.
Also shown by a dashed line is the equal sharing of the excitation energy as well as the band
corresponding to the case for equal temperature thermal equilibrium for PLF and TLF
for which the sharing of the excitation energy $E^{*}_{PLF}/E^{*}_{Total}=A_{PLF}/A_{Total}$~\cite{planeta1989}.
The vertical width of the band reflects the distribution of $A_{PLF}$. 
We observe that for low $E_\mathrm{loss}$ values the partition of the excitation energy
is closer to the equal sharing line.  
However, as the energy loss increases, the repartition of excitation energy 
gets closet to the thermal equilibrium limit but rarely reach there.
It is satisfactory to see that these fully microscopic calculations, 
with no parameter adjusted on reaction properties, are able to affirm previous experimental observations~\cite{awes1984,vandenbosch1984}. 

\section{Summary and discussion}
\label{sec:summary}

Time-dependent Hartree-Fock (TDHF) method in full 3D is employed to calculate deep-inelastic reactions of 
$^{78}$Kr+$^{208}$Pb and $^{92}$Kr+$^{208}$Pb systems. The impact parameter and energy-loss
dependence of relevant observables are calculated. In addition, density constrained TDHF method is used
to compute excitation energies of the primary fragments. The statistical deexcitation code GEMINI is utilized
to examine the final reaction products.

We find a smooth dependence of the energy loss, $E_\mathrm{loss}$, on the impact parameter for both
systems. On the other hand the transfer properties for low $E_\mathrm{loss}$ values are very different
for the two systems but become similar in the higher $E_\mathrm{loss}$ regime.
The impact parameter dependence of transfer shows more structure emanating from shell effects and
orientation of the deformed projectile.

A charge equilibration process is observed when the nuclei have an initial $(N/Z)$ asymmetry, 
with an increased $(N/Z)$ equilibration as the energy damping is increased. 
However, even fully damped collisions usually do not lead to identical $(N/Z)$ values in the fragment. 
This is because the $(N/Z)$ content of the fragment is only an approximate proxy for the chemical potential. 
Nevertheless, the evolution of the $(N/Z)$ values of the fragments as a function of contact time  can be used 
to investigate the charge equilibration process. 
Experimentally, contact times are not a direct observable, but can be reconstructed by comparison with theoretical predictions of the fragment properties (mass, charge, scattering angle and kinetic energy). 
The present TDHF calculations indicate a mean lifetime of charge equilibration of $\sim 0.5$~zs, of the same order than mean lifetime obtained from experimental data at Fermi energies. 

The fully microscopic TDHF theory has shown itself to be rich in
nuclear phenomena and continues to stimulate our understanding of nuclear dynamics.
The time-dependent mean-field studies seem to show that the dynamic evolution
builds up correlations that are not present in the static theory.
While, there is evidence that one-body dissipation can properly account for the transport phenomena
seen in these reactions further experiments are needed to test this conclusion.

\begin{acknowledgments}
This work has been supported by the
U.S. Department of Energy under grant No. DE-SC0013847,
by the Australian Research Council Grants No. FT120100760 and DP160101254, and
by the National Natural Science Foundation of China under Grant
No. 11575044 and the China Scholarship Council (File No.
201606095006).
\end{acknowledgments}
\bibliography{VU_bibtex_master.bib}

\end{document}